\documentclass[submission,copyright,creativecommons]{eptcs}
\providecommand{\event}{PLACES 2025} % Name of the event you are submitting to
\usepackage{iftex}

\ifpdf
  \usepackage{underscore}         % Only needed if you use pdflatex.
  \usepackage[T1]{fontenc}        % Recommended with pdflatex
\else
  \usepackage{breakurl}           % Not needed if you use pdflatex only.
\fi

\usepackage{listings}             % for typesetting program code
\usepackage{multicol}             % for typesetting program code
\usepackage{graphicx}
\usepackage{amsmath}
\usepackage{wrapfig}
\lstdefinelanguage{Lime}
{
  morekeywords={
    class, inherit, init, method, action, when, while, if, else, then, new, do, od, int, bool, return, var, const, for, to, not, elif, and, or, array, of, await, procedure 
  },
  sensitive=false, % keywords are not case-sensitive
  morecomment=[l]{//}, % l is for line comment
  morecomment=[s]{/*}{*/}, % s is for start and end delimiter
  morestring=[b]" % defines that strings are enclosed in double quotes
}
\newcommand{\choice}{\mathbin{[\hspace{-0.04em}]}}
\newcommand{\semi}{\mathbin{;}}

\title{An Efficient Implementation of Guard-Based Synchronization for an Object-Oriented Programming Language}
\author{
Shucai Yao
\institute{Huawai Technologies Canada \\
Markham, Ontario, Canada\thanks{Work performed while affiliated with McMaster University.}}
\email{yao2001626@gmail.com}
\and
Emil Sekerinski
\institute{McMaster University \\ Hamilton, Ontario, Canada}
\email{emil@mcmaster.ca}
}
\def\titlerunning{Efficient Implementation of Guard-Based Synchronization}
\def\authorrunning{S. Yao, E. Sekerinski}
\begin{document}
\maketitle

\begin{abstract}
In the shared variable model of concurrency, guarded atomic actions restrict the possible interference between processes by regions of atomic execution. The guard specifies the condition for entering an atomic region. That is a convenient model for the specification and verification of concurrent programs, but it has eschewed efficient execution so far.
%While guarded atomic actions are found verification tools, programming languages eschew them in favour of semaphores, monitors, message passing, etc. for synchronization and communication, on the grounds of efficiency.
This article shows how guarded atomic actions, when attached to objects, can be implemented highly efficiently using a combination of coroutines, operating-system worker threads, and dedicated management of object queues and stacks. The efficiency of an experimental language, Lime, is shown to compare favourably with that of C/Pthreads, Go, Erlang, Java, and Haskell on synthetic benchmarks.
\end{abstract}

\section{Introduction}

For concurrency based on \emph{shared variables}, there is a long history of specifying synchronization and atomicity by \emph{atomic guarded commands} (\emph{atomic actions}), e.g.~in conditional critical regions~\cite{hoare1972towards}, the Owicki-Gries theory~\cite{owicki1976axiomatic}, Unity~\cite{misra1988parallel}, action systems~\cite{back1989decentralization},  TLA \cite{lamport1994temporal}, Seuss \cite{misra2001discipline}, as well as in model checkers for concurrent programs. Atomic guarded commands are also used in verification tools, e.g. Event-B~\cite{abrialModelingEventBSystem2010} and CIVL~\cite{10.1145/2807591.2807635}. % $\$when (expr) stmt;$
While an implementation of conditional critical regions by software transactional memory~\cite{harris2003language} was proposed, Occam~\cite{inmos1984occam}, Ada~\cite{ichbiah1979rationale, brosgol1997comparison}, and Go~\cite{pike2009go} allow limited forms of atomic guarded commands: none of these supports guarded commands $g \to S$ where the guard $g$ being true initiates the execution of $S$. The common wisdom is that guarded commands cannot be implemented efficiently: ``the price that must be paid for this automatic scheme is performance''~\cite{briot1998concurrency}. Instead, mainstream languages offer semaphores~\cite{dijkstra1962over,dijkstra1968structure}, monitors~\cite{hoare1974monitors}, and variations thereof.

Given the suitability---and today's ubiquity---of object-oriented languages for modelling program domains, the notion that objects are naturally ``units of concurrency'' emerged early on~\cite{birtwisleSimulaBegin1975,ishikawa1984design}. The \emph{actor model}~\cite{agha1985actors} with asynchronous message passing stems from early work on (rule-based) AI systems~\cite{hewitt1971procedural}. Erlang is the first inherently concurrent programming language that implements the actor model~\cite{armstrong1993concurrent}. The actor model has since then been used by Scala and other programming languages. The Eiffel programming language differs from that by using method calls for synchronous communication and re-interprets preconditions as method guards~\cite{west2015efficient}. Yet another form of concurrent objects is proposed in~\cite{faes2018concurrency}.

CSP extends the notion of a guarded command by allowing the guard to contain synchronous communication with other processes of a program~\cite{hoare1978communicating}. CSP influenced the designs of Occam~\cite{inmos1984occam} and Go~\cite{pike2009go}. In these languages, the structure of programs is dominated by processes (called goroutines in Go) and the object structure is de-emphasized. Processes are created explicitly, rather than being implicitly started by guarded commands of the form $g \to S$.

Work on the correctness and refinement of action systems led to natural object-oriented extensions~\cite{bonsangue1998approach,buchi2000foundation,sekerinski2002concurrent}: objects communicate by method calls, synchronize by guarded methods, and have atomic actions that specify concurrent execution. As actions are atomic, execution would need backtracking if an object with a blocking method is called: if an action with body $S \mathbin; x.m() \mathbin T$ is called and method $m$ blocks, the effect of $S$ has to be reversed as the action must either be executed to completion or not executed. If $S$ contains method calls, the effect of that call has to be reversed. While the model is simple, no efficient implementations exist.

% Futures: The concept of \textit{futures}, first introduced by Baker~\cite{baker1977incremental}, was used for synchronizations in concurrent programming languages. A somewhat similar concept, called \textit{promise} was proposed by Friedman~\cite{friedman1976impact}.

% Limitations of threads: lightweight threads, such as fibers~\cite{shankar2003implementing}, green threads~\cite{sunsoft2}. 

% Coroutines: Knuth \cite{knuth1997art},  C\#~\cite{asynchronousmode19microsoft} and ECMAScript~\cite{asyncfunction19Mozilla}.

This work explores how guarded commands can be implemented highly efficiently when viewing objects as the ``unit of concurrency''. Object-oriented action systems are taken as the basis and modified to allow execution without backtracking. %The implementation relies on the tight integration of compiler and a dedicated run-time system with user-level coroutines. As high efficiency cannot be achieved by extensions of existing languages, an experimental language, Lime, was designed.

The following section introduces our experimental language, Lime, through examples and defines it in terms of guarded commands with parallel composition and atomicity brackets. Section~3 discusses the scheme for guard evaluation, the implementation with cooperative scheduling of user-level coroutines, and the runtime system with object queues local to worker threads and global queues. Section~4 presents three synthetic benchmarks with fine-grained concurrency.

%The current implementation of Lime uses cooperative scheduling, user-level coroutines, a runtime system with a local object queue for each worker thread and a global queue, a lock-free implementation of the local object queue, and a segmented stack mechanism. The implementation performs better on three synthetic benchmarks with fine-grained high concurrency than C/Pthreads, Go, Erlang, Java, and Haskell.

% First efficient implementation of guard-based synchronizations for Lime, which has a better performance than concurrent programming languages such as Erlang and Go, in fine-grained, highly concurrent benchmarks
% 2. First implementation of cooperative scheduling for Lime
% 3. First implementation of user-level coroutines for Lime
% 4. First implementation of Lime runtime system which maintains a local object queue for the worker thread and a global object queue
% 5. First lock-free implementation of the local object queue for Lime
% 6. Evaluation of segmented stack mechanisms for highly concurrent objects

\section{An Action-based Object-oriented Programming Language}

\begin{wrapfigure}{r}{.35\linewidth}
\vspace{-16pt}
\begin{lstlisting}
class Doubler 
    var x: int
    init()
        this.x := 0
    method store(u: int)
        this.x := 2 * u 
    method retrieve(): int
        return this.x
class DelayedDoubler 
    var y: int
    var d: bool
    init()
        this.y, this.d := 0, true
    method store(u: int)
        this.y, this.d := u, false
    method retrieve() : int
        when this.d do
            return this.y
    action double 
        when not this.d do 
            this.y, this.d := 2 * y, true 
\end{lstlisting}
\caption{Delayed doubler in Lime}
\label{code:ddlime}
\end{wrapfigure}

Lime uses indentation for bracketing. The guarded command $g \to S$ is written as \lstinline|when g do S|, where $g$ is a Boolean expression and $S$ is a statement. Methods and actions can be guarded. When a method is called and its guard is false, the call is suspended; it can be resumed when the guard becomes true. Actions are repeatedly executed by selection an action with a true guard nondeterministically. Actions have a name, but cannot be called. Only one method or action in an object can execute at a time, but multiple objects can execute concurrently.

The class \emph{Doubler} in Figure~\ref{code:ddlime} allows an integer to be stored and its double value to be retrieved~\cite{cui2009experimental,sekerinski2002concurrent}. The class \emph{DelayedDoubler} performs the same functionality but doubles ``in the background'': method \emph{store} sets field $d$ to \emph{false}, which \emph{blocks} calls to \emph{retrieve} until the action \emph{double} performs doubling and sets $d$ to \emph{false}. Thus, a call to \emph{store} can return quickly. This example is representative of calls that enable a background activity, like storing data in files, sending data over a network, or requesting remote data. The methods and actions in an object are executed \emph{atomically up to method calls}. Since \emph{Doubler} and \emph{DelayedDoubler} do not contain method calls, all methods and actions are executed atomically. Here, $x \bmod 2 = 0$ is an invariant of \emph{Doubler} and \emph{DelayedDoubler} refines \emph{Doubler} through the relation $d \Rightarrow y = 2 * u$.

In Lime, only an object's own fields can be accessed; thus, we leave out \textit{this} in subsequent examples. Method and action guards must be only over the fields of an object.

\begin{figure}\center
 \includegraphics[width=.7\linewidth]{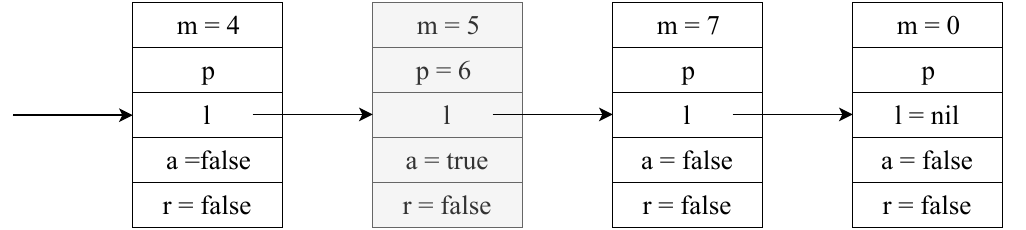}
 \caption{Possible state of a priority queue after adding 4, 5, 7, 6}
 \label{fig:PDdemo}
\end{figure}

\begin{figure}[tb]
% \begin{wrapfigure}{r}{.55\linewidth}
\begin{lstlisting}[multicols=2]
class PriorityQueue 
    var m,p: int
    var l: PriorityQueue
    var a,r: bool
    init() 
        l, a, r, m := nil, false, false, 0 
    method empty() : bool
        when not r do 
            return l = nil 
    method add(e: int) 
        when not a and not r do
            if l = nil then 
                m, l := e, new PriorityQueue()    
            else 
                p, a := e, true 
    method remove() : int
        when not a and not r do 
            r := true 
            return m 
    action doAdd 
        when a do 
            if m < p then 
                l.add(p) 
            else 
                l.add(m) 
                m := p 
            a := false 
    action doRemove 
        when r do 
            if l = nil then  
                r := false 
                return
            elif l.empty() then
                l := nil
            else
                m := l.remove()
            r := false 
\end{lstlisting}
\caption{Priority Queue in Lime}
\label{code:pqlime}
% \end{wrapfigure}
\end{figure}

Figures \ref{fig:PDdemo} and \ref{code:pqlime} show a priority queue adapted from~\cite{Sekerinski03SimpleCOOP}. Method \textit{add}(\textit{e}) stores a positive integer \textit{e}, method \textit{remove} removes the least integer stored, and method \textit{empty} tests whether the priority queue is empty. Elements are stored in field \textit{m} in ascending order (duplicates are allowed). The priority queue starts with a sentinel node ($m = 0$). Field \textit{l} points to the next node or is \textit{nil}. An element is added to the priority queue by either storing it in the current node if it is the last one (and creating a new last node) or by depositing it in field \textit{p} of the current node and enabling action \textit{doAdd} that will move either the new element or the element of the current node one position down. The minimal element is removed by returning the element of the current node and enabling action \textit{doRemove} that will move the element of the next node one position up or set \textit{l} to \textit{nil} if the node becomes the last. In principle, all nodes of a priority queue can work concurrently.

Like the delayed doubler, the priority queue implements an \textit{early return}, which is cumbersome to express with semaphores or monitors but more general than futures. Actions \textit{doAdd} and \textit{doRemove} contain method calls; as actions (and methods) are atomic only up to method calls, when \textit{l.add}() is called in \textit{doAdd}, atomicity stops and in principle methods \textit{empty} and \textit{add} can be called. This is prevented as field \textit{a} remains \textit{true} and is set to \textit{false} only at the end of \textit{doAdd}. An invariant is $\neg(a \land r)$.

\begin{figure}[tb]
\begin{lstlisting}[multicols=2]
class Node
    var key, p: int
    var left, right: Node
    var a: bool
    init(x: int)
        key, left, right, a := x, nil, nil, false
    method add(x: int)
        when not a do
            if left != nil then a, p := true, x
            elif x < key then
                left,right, key := new Node(x), new Node(key), x
            elif x > key then
                left,right := new Node(key), new Node(x)
    method has(x: int): bool
        when not a do
            if left = nil then return x = key
            elif x <= key then return left.has(x)
            else return right.has(x)
    action addToChild
        when a do
            if p <= key then left.add(p)
            else right.add(p)
            a := false
\end{lstlisting}
\caption{Leaf-oriented Tree in Lime}
\label{code:lotlime}
\end{figure}

The leaf-oriented tree in Fig.~\ref{code:lotlime} implements concurrent insertion in a set. It is adapted from~\cite{Sekerinski03SimpleCOOP}. The internal nodes contain only guides; the elements are stored in the leaves. Insertion either creates two new leaves, one with the original element and one with the element to be inserted, or deposits an element in an internal node. Each node has an action that eventually moves the deposited element one level closer to its final position. This action must hold a lock only on the current node and one of its children. Thus, insertions can proceed in parallel in different parts of the tree. The methods \textit{add} and \textit{has} are guarded to prevent possible overtaking.

\begin{figure}[tb]
\begin{lstlisting}[multicols=2]
class Reducer
    var index: int
    var next: Reducer 
    var a1, a2: bool
    var e1, e2: int
    init(i: int, r: Reducer)
        index, a1, a2, next := i, false, false, r
    method reduce1(x: int)
        when not a1 do 
            e1, a1 := x, true
    method reduce2(x: int)
        when not a2 do 
            e2, a2 := x, true
    action doReduce
        when a1 and a2 do 
            if index = 1 then
                print(e1 + e2)
                e1, e2 := 0, 0
            elif index % 2 = 0 then 
                next.reduce1(e1 + e2)
            else
                next.reduce2(e1 + e2)
            a1, a2 := false, false 
class Mapper
    var  next: Reducer
    var  a: bool 
    var  e, index: int 
    init(i: int, r: Reducer)
        index, a, next := i, false, r
    method map(n: int)
        when not a do
            e, a := n, true
    action doMap
        when a do 
            if index % 2 = 0 then 
                next.reduce1(e * e)
            else 
                next.reduce2(e * e)
            a := false
\end{lstlisting}
\caption{Map-reduce in Lime}
\label{code:mrlime}
\end{figure}

The final example is the map-reduce programming model: a \textit{map} function is applied to each input element, and the results are combined with a \textit{reduce} function to a single result. The classes in Fig.~\ref{code:mrlime} allow mapping and reducing to proceed concurrently. The execution time consists of the time for communication and the computation of \emph{map} and \emph{reduce}. Since our goal is to measure the communication time, the computational is kept small: \emph{map} squares an element and \emph{reduce} adds two elements. The main program creates one \textit{Mapper} object for each input element and links the \textit{Reducer} objects as a tree.

Lime is defined in terms of \emph{guarded commands} with \emph{parallel composition} ($\ldots\parallel\ldots$) and \emph{atomicity brackets} ($\langle\ldots\rangle$), for which verification rules are well-established, e.g.~\cite{andrews1991concurrent}. The construct $\langle g \to S \rangle$ evaluates $g$ and executes $S$ atomically. For every class $C$, a variable, also called $C$, is introduced with the set of objects of $C$. For every field $v$ of $C$, a variable \textit{C_v} for the value of $v$ for each $C$ object is introduced; a boolean field \textit{lock} is added. Procedure \textit{C_new} creates a new object of class $C$ and executes its initialization. For each method $m$ of $C$, a procedure \textit{C_m} is introduced that takes an additional \textit{this} parameter. For each (parameterless) action $a$ of class $C$, a procedure \textit{C_a} is introduced with a \textit{this} parameter. The set \textit{Ref} of object references includes the value \textit{nil}, Fig.~\ref{fig:classdef}. 

\begin{figure}[tb]
\begin{minipage}{.32\columnwidth}
\begin{lstlisting}
class C
  var v: V
  init ()
    I
  method m$_0$(u$_0$: U$_0$) $\to$ (w$_0$: W$_0$)
    when g$_0$ do M$_0$
  method m$_1$(u$_1$: U$_1$) $\to$ (w$_1$: W$_1$)
    when g$_1$ do M$_1$
  ...
  action a$_0$
    when h$_0$ do A$_0$
  action a$_1$
    when h$_1$ do A$_1$
  ...



    
\end{lstlisting}
\end{minipage}
\begin{minipage}{.68\columnwidth}
\begin{lstlisting}
var C: set(Ref) := {}
var C_lock: Ref $\to$ bool
var C_v: Ref $\to$ V
procedure $\it C\_new() \to$ (this: Ref)
  $\it\langle this :\not\in C \cup \{nil\} \semi C := C \cup \{this\} ; this.lock := true \rangle;~I;~ this.lock := false$
procedure C_m$_0$(u$_0$: U$_0$, this: Ref) $\to$ (w$_0$: W$_0$)
  $\it \langle g_{\rm 0} \land \neg this.lock \to this.lock := true \rangle;~M_{\rm 0};~this.lock := false$
procedure C_m$_1$(u$_1$: U$_1$, this: Ref) $\to$ (w$_1$: W$_1$)
  $\it\langle g_{\rm 1} \land \neg this.lock \to this.lock := true \rangle;~M_{\rm 1};~this.lock := false$
...
procedure C_$a_0$(this: Ref)
  $\langle$ $h_0 \land \neg$ this.lock $\to$ this.lock := true $\rangle;~A_ 0;~$ this.lock := false
procedure C_$a_1$(this: Ref)
  $\langle$ $h_1 \land \neg$ this.lock $\to$ this.lock := true $\rangle;~A_ 1;~$ this.lock := false
...
\end{lstlisting}
\end{minipage}
% \begin{minipage}{.68\columnwidth}
% \begin{lstlisting}
% var C: set(Ref) := {}
% var C.lock: Ref $\to$ bool
% var C.v: Ref $\to$ V
% procedure $\it C\_new() \to$ (this: Ref)
%   $\it\langle this :\not\in C \cup \{nil\} \semi C := C \cup \{this\} ; this.lock := true \rangle;~I;~ this.lock := false$
% procedure C_m$_0$(u$_0$: U$_0, this: Ref$) $\to$ (w$_0$: W$_0$)
%   $\it \langle g_{\rm 0} \land \neg this.lock \to this.lock := true \rangle;~M_{\rm 0};~this.lock := false$
% procedure C_m$_1$(this: Ref, u$_1$: U$_1$) $\to$ (w$_1$: W$_1$)
%   $\it\langle g_{\rm 1} \land \neg this.lock \to this.lock := true \rangle;~M_{\rm 1};~this.lock := false$
% procedure C_action(this: Ref)
%   do $\it \langle h_{\rm 0} \land \neg this.lock \to this.lock := true \rangle;~A_{\rm 0};~this.lock := false$
%   $\it ~\choice~~\langle h_{\rm 1} \land \neg this.lock \to this.lock := true \rangle;~A_{\rm 1};~this.lock := false$
% \end{lstlisting}
% \end{minipage}
\caption{Definition of a class in terms of guarded commands. A method or action guard that is \textit{true} can be left out.}
\label{fig:classdef}
\end{figure}

Accessing field $v$ within the methods and actions of a class stands for \textit{this.v}. Suppose that $o$ is declared of class $C$. In general, $o.x$ stands for $C\_v(o)$. Calling $o.m$ involves releasing the lock to the current object, calling $C\_m$, and locking the current object upon return. This avoids the need for backtracking and makes actions and methods atomic up to method calls. Creating a new object of class $C$ involves calling $\it C\_new$:
\begin{align*}
& o.v                   && = && C\_v(o) \\
& x := o.m(e)           && = && {\it this}.{\it lock} := {\it false} \semi x := C\_m(o, e) \semi \langle \neg {\it this}.{\it lock} \to {\it this}.{\it lock} := {\it true} \rangle \\
& o := \textbf{new}~C(e) && = && o := C\_{\it new}(e)
\end{align*}

% Suppose a program consists of classes $C, D$, and \textit{Start}. Its behaviour is defined as executing all actions and the initialization of \textit{Start} in parallel: 
% \begin{align*}
% (\parallel o \in C \to C\_{\it action}(o)) \parallel
% (\parallel o \in D \to D\_{\it action}(o)) \parallel
% (\textbf{var}~s: {\it Start}; s:=\textbf{new}~Start())
% \end{align*}

\noindent Suppose a program consists of classes $C_0, C_1, \ldots$ and class \textit{Start} is among those. The program's behaviour is defined as executing all \emph{enabled} actions, i.e. with a true guard, of all objects and the initialization of \textit{Start} in parallel. Initially, there are no objects:
\begin{align*}
(\parallel o \in C_0 \to C_0\_a_0(o) \choice C_0\_a_1(o) \choice \ldots) \parallel \ldots \parallel
(\textbf{var}~s: {\it Start}; s:=\textbf{new}~Start())
\end{align*}

\section{Implementation}

The Lime runtime maps $M$ \emph{active objects}, i.e., objects with actions, to $N$ worker (operating system) threads, where $N$ is typically less than the number of CPU cores. For each active object with a running action, a coroutine with its own stack is created. The stack is \emph{segmented}. As most actions do not make deep recursive calls, the stack is initially small, with only 4 KB, and grows as needed. For this, extra code on method calls is inserted that checks if a stack overflow is about to happen. As the overhead of these checks can accumulate, the stack calling convention is modified to minimize the impact~\cite{Yao20GuardBasedSynchronization}.

The coroutines are scheduled cooperatively. The compiler takes a Lime source file with a class \textit{Start} and generates two files, an x86 assembly file with the code for guard evaluation and context switches to the scheduler, and a C file with method and action bodies. LLVM is used to compile and optimize the C code. LLVM does not offer hooks for coroutine switching as needed here, so it is only used for method and action bodies. The scheduler is part of the runtime system and is linked with the generated assembly files and compiled C files. The EBP register is reserved for \textit{this}, the pointer to the current object. When a worker thread switches the context from one object to another, only three registers, EBP, ESP (stack pointer), and EIP (instruction pointer), need to be saved and restored; no other registers are in use at the time of context switches. This makes switching faster than preemptive scheduling, where all registers typically need to be saved and restored.

Following the formal definition, each object has a hidden boolean field, \textit{lock}, that is initially \textit{false}. The \emph{originator} is the object with the action that initiated a computation, or the \emph{Start} object. At each method call, the originator is passed as an additional parameter. The call $o.m()$ first locks $o$ and then evaluates the guard. If the object is locked or the guard is false, control is transferred to the scheduler. Otherwise, the body is executed and the object is placed in \emph{runQ}, if the object has actions, Fig.~\ref{fig:methodactiontrans}. The call $x := o.m(e)$ is translated to:
\begin{lstlisting}
    unlock(this.lock) ; x := C_m(e, o, originator) ; lock(this.lock)
\end{lstlisting}

\begin{figure}[tb]
\begin{minipage}{.55\columnwidth}
\begin{lstlisting}
procedure C_m(u: U this: Ref, originator: Ref) $\to$ (w: W)
    while true do 
        if lock(this.lock) then
            if g then 
                M 
                runQ.put(this)
                unlock(this.lock) 
                return
            else 
                unlock(this.lock) 
        switch_to_sched(originator)
\end{lstlisting}
\end{minipage}
\begin{minipage}{.45\columnwidth}
\begin{lstlisting}
procedure C_a(this: Ref)
    const originator = this
    while true do 
        if lock(this.lock) then 
            if h1 then 
                A 
                unlock(this.lock) 
            else 
                unlock(this.lock) 
        switch_to_sched(originator)
    




\end{lstlisting}
\end{minipage}
\caption{Translation schema for method $m$ (left) and action $a$ (right) of class $C$ of Fig.~\ref{fig:classdef}.}
\label{fig:methodactiontrans}
\end{figure}

\begin{figure}[tb]\center
\begin{minipage}[t]{0.49\linewidth}
\includegraphics[width=1.05\linewidth]{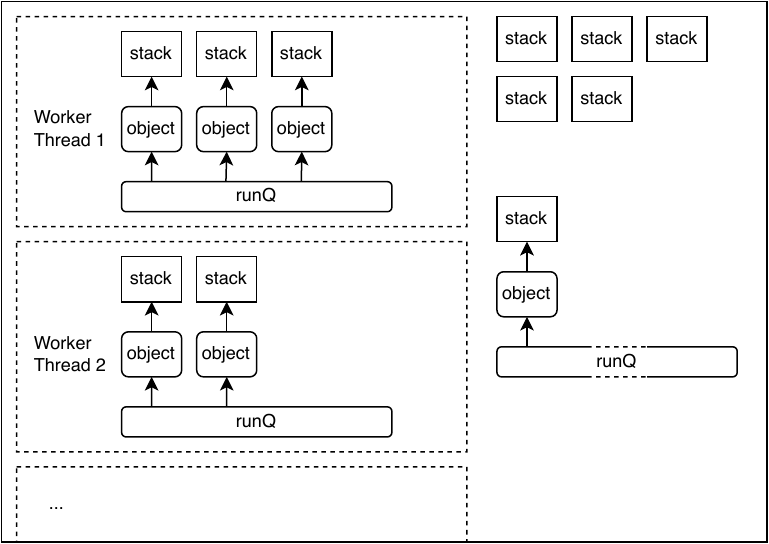}
\caption{Lime runtime structure; the thread-local queues are of fixed size and the global queue can grow as needed.}
 \label{fig:runtime}
\end{minipage}
% \end{figure}
% \begin{figure}[tb]
\begin{minipage}[t]{0.49\linewidth}
\centering
\includegraphics[width=1.05\linewidth]{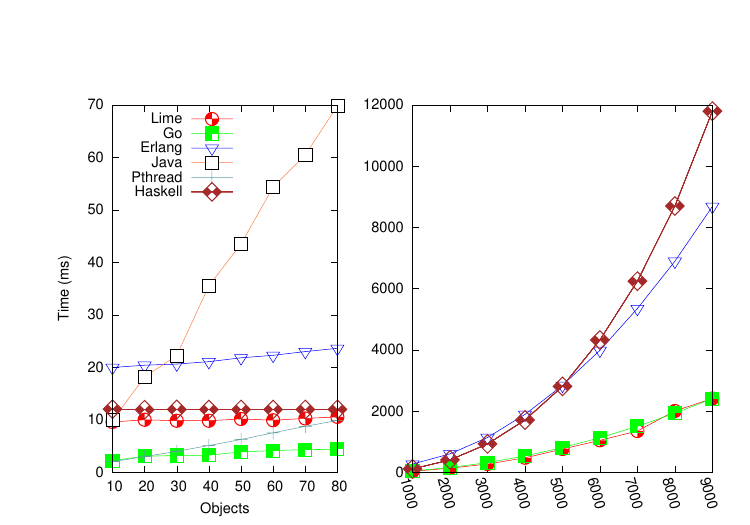}
\caption{Priority queue timing results}
\label{fig:PQresult}
\end{minipage}
\end{figure}

\noindent Each worker thread has its own \emph{runQ} queue with objects that may contain an enabled action or have been suspended. Periodically, the worker threads evaluate the guards of the actions of objects in \emph{runQ} and execute them.
%Suspended originators are placed in the \emph{suspendQ} queue. Only objects in \emph{suspendQ} have a stack allocated.
Since actions can start and terminate frequently, stacks are preallocated and shared among all worker threads, Fig.~\ref{fig:runtime}.

When a worker thread is initialized or runs out of objects, it can fetch objects from the global queue. If that is empty, it can \emph{steal} an object from another worker. For this, local queues are implemented as double-ended queues with lock-free synchronization.

The Go and Erlang implementations use the $M:N$ threading model. While Erlang relies on asynchronous message passing, Go supports synchronous and asynchronous message passing, with a stack allocated for each goroutine~\cite{lockfreequeue2025go}. Go’s runtime utilizes local lock-free queues for each worker thread. When a local queue is empty, work stealing is employed to retrieve tasks from other workers. The Erlang runtime system, similarly, uses a work-stealing scheduler to manage actors and distribute the workload evenly, also utilizing local lock-free queues~\cite{lockfreequeue2025erlang}. The key difference is in the use of guards versus channels.

\section{Timing Results}

\begin{figure}[tb]
\begin{minipage}[t]{0.5\linewidth}
\centering
\includegraphics[width=1.05\linewidth]{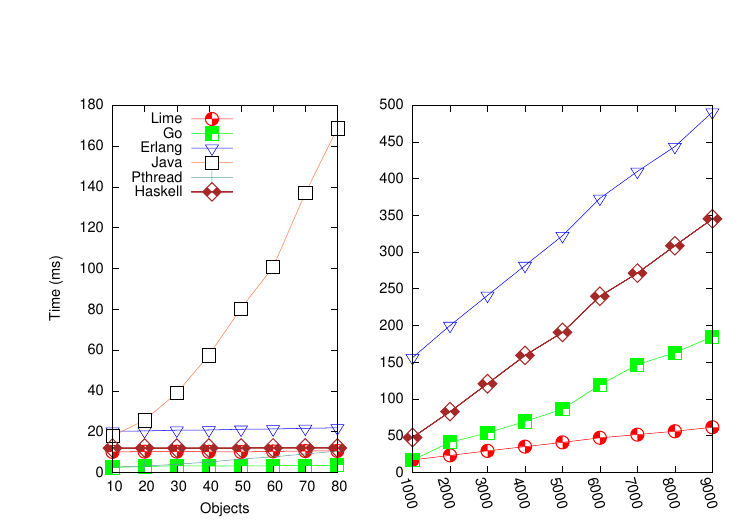}
\caption{Leaf-oriented tree timing results}
\label{fig:LOTresult}
\end{minipage}
% \end{figure}
% \begin{figure}[tb]
\begin{minipage}[t]{0.5\linewidth}
\centering
\includegraphics[width=1.05\linewidth]{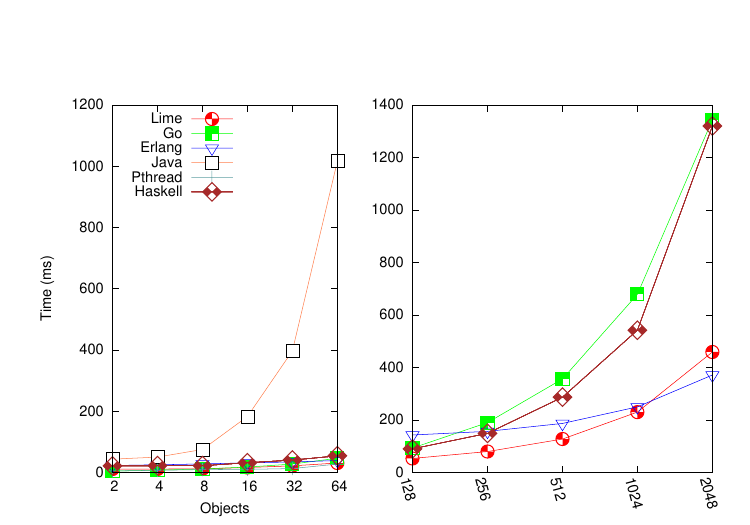}
\caption{Map-reduce timing results}
\label{fig:MRresult}
\end{minipage}
\end{figure}

To isolate synchronization and communication overhead from other computations, three programs with little computation but representing different concurrency patterns are selected: \emph{priority queue}, a linear structure, \emph{map-reduce}, a tree structure with the computation starting at the leaves, and \emph{leaf-oriented tree}, a tree structure with the computation starting at the root.

Lime is compared with Java (OpenJDK 19.0.2), C/Pthread (GCC 14.4.0), Erlang (Erlang/OTP 24), Go (golang 1.18.1), Haskell (ghc 8.8.4). The complete listings for these programs can be obtained from the project's GitLab repository (\url{https://gitlab.cas.mcmaster.ca/yaos4/thesis_code.git}). The experiments were run on AMD Ryzen Threadripper 3990X 64-Core Processor (2.2 - 4.4GHz). All measurements were performed with Ubuntu (22.04 LTS) in single-user mode. The execution time is measured by the Unix \emph{time} command, and each timing measurement is repeated thirty times. The results reported here are the average with a confidence interval of 95\%. As the difference between the maximum and minimum values is small enough, only the average value is reported. For Erlang, because there is a constant overhead of around 1000 ms for starting and stopping the virtual machine, that time is subtracted. There is a shorter startup time for Java. For Erlang and Java, the tests are repeated 10 times to amortize this overhead. All implementations use the same sequence of pseudo-random numbers.

The timing results in Figs.~\ref{fig:PQresult},~\ref{fig:LOTresult}, and~\ref{fig:MRresult} are split into plots with ``small'' and ``large'' number of objects. The plots show that the lightweight thread implementations of Go, Erlang, Haskell, and Lime outperform the heavyweight thread implementations of Java and Pthread; the times for those are not shown in the right-hand plots. Secondly, the coroutine thread implementations of Go and Lime generally outperform other lightweight implementations.

In the priority queue, the head node is the bottleneck, the second node is the second most busy node, etc. This tests how well the threads select node objects to work on.

In the leaf-oriented tree, the root object is also the bottleneck. Suppose that the tree is perfectly balanced with 10,000 nodes. Although, in principle, 5,000 node objects can execute concurrently, the real opportunity for concurrency is low: if each node spends the same time passing the data, only 0.14\% (14/10,000) of the nodes would execute concurrently since the approximate depth of the tree is 14. This tests how well the runtime system performs if there are many objects but only a few can execute.

In map-reduce, the inputs are the integers $0$ to $num - 1$. The computation is repeated \textit{repeat} times to ``fill the pipeline''. In this example, there is an abundance of possible concurrency; it tests how well the runtime systems exploit that.

A more thorough discussion of the timing results can be found in~\cite{Yao20GuardBasedSynchronization}.

\section{Conclusions}

This research started as an experiment to evaluate what language constraints are needed and which implementation techniques are suitable to execute atomic guarded commands. The language, Lime, incorporates objects as the ``unit'' of concurrency, thus unifying the concepts of processes and objects. The results are favourable compared to well-established implementations on a small set of carefully selected benchmarks. In the language, the efficiency is achieved by (1) weakening the total atomicity of actions to atomicity only up to (potentially blocking) method calls, thus avoiding the need for backtracking, and (2) restricting guards to be only over the fields of an object, necessitating that guards in an object are reevaluated only after a call to the object or a method call from within that object. In the implementation, the efficiency is achieved by (1) allocating for each executing object a small stack that can grow as needed, (2) implementing each object as a user-level coroutine with fast cooperative scheduling (requiring only three registers to be saved and restored when switching stacks), (3) employing at most as many worker threads as there are cores, (4) using a combination of (lock-free) local and global queues for load balancing and work stealing, and (5) modifying the procedure call to allow an efficient detection of stack overflow. The thesis~\cite{Yao20GuardBasedSynchronization} discusses alternative implementations without lock-free queues and using two local queues (one for objects with a stack and one for objects without a stack). Further experiments with ``real'' programs and more complex guards are needed to determine how well a Lime-like language works in practice.

\paragraph{Acknowledgements.} The reviewers made useful suggestions, for which the authors are grateful.

\bibliographystyle{eptcs}
\bibliography{references}

\end{document}